\documentclass{JHEP3}
\usepackage{amsmath}
\usepackage{amssymb}
\usepackage{graphicx}
\usepackage{cite}
\usepackage{tabularx}
\title{QCD dynamics in a constant chromomagnetic field}

\author{Paolo Cea \\ Universit\`a di Bari  \& INFN - Bari \\
via Amendola 173 - 70126 Bari - Italy \\
\email{paolo.cea@ba.infn.it}}

\author{Leonardo Cosmai  \\ INFN - Bari \\
via Amendola 173 - 70126 Bari - Italy \\
\email{leonardo.cosmai@ba.infn.it}}

\author{Massimo D'Elia  \\  Universit\`a di Genova \& INFN - Genova \\
via Dodecaneso 33 - 16146 Genova - Italy \\
\email{massimo.delia@ge.infn.it} }

\received{\today}

\abstract{We investigate the phase transition in full QCD with two
flavors of staggered fermions in presence of a constant abelian
chromomagnetic field. We find that the critical temperature depends
on the strength of the chromomagnetic field and that the deconfined
phase extends to very low temperatures for strong enough fields. As
in the case of zero external field, a single transition is detected,
within statistical uncertainties, where both deconfinement and chiral
symmetry restoration take place. We also find that the chiral
condensate increases with the strength of the chromomagnetic field.}

\keywords{Lattice QCD, Confinement}

\preprint{
BARI-TH 2007/568 \\
GEF-TH-15-07}

\begin{document}

\newcommand{\be}{\begin{equation}}
\newcommand{\ee}{\end{equation}}

\section{Introduction}
\label{Introduction}
In previous studies~\cite{Cea:2002wx,Cea:2005td}  on the vacuum dynamics of
pure non-abelian gauge theories we found that the deconfinement
temperature depends on the strength of an external abelian
chromomagnetic field. In particular we ascertained that the deconfinement
temperature decreases when the strength of the applied field is
increased and eventually goes to zero. It is not difficult to see here an
analogy with the reversible Meissner effect in the case of ordinary
superconductors (where the system goes to normal even at zero temperature
if the magnetic field is strong enough)
and  therefore we referred to it as "vacuum color
Meissner effect". We have also verified that the same effect
is not present in the case of abelian gauge theories, so that
it seems to be directly linked to the non-abelian nature of
the gauge group.

The dependence of the deconfinement temperature on
applied external fields is surely linked to the dynamics
underlying color confinement, therefore in our opinion,
apart from possible phenomenological implications,
such an effect could shed light on confinement/deconfinement
mechanisms.

On these basis we believe that it is important to
test if the effect continues to hold and how it qualitatively
changes when switching on fermionic degrees of freedom.
One important aim of the present work  is therefore to investigate
the dependence of the deconfinement temperature
on the strength of an external abelian chromomagnetic field
in the case of  full QCD with two flavors.

A second important and relevant issue regards the relation
between deconfinement and chiral symmetry restoration.
As it is well known, the two phenomena appear to be coincident
in ordinary QCD, while they are not so in different theories
(like QCD with adjoint fermions~\cite{Karsch:1998qj,Engels:2005rr,Lacagnina:2006sk}).
A simple explanation of this fact is not yet known
and may be strictly linked to the very dynamics of color confinement.
An important contribution towards a clear understanding of
this phenomenon could be to study whether it is stable
against the variation of external parameters: both
theoretical and numerical
studies~\cite{McLerran:2007qj,Hands:2006ve,Alles:2006ea,Conradi:2007kr}
have been performed in that sense for the case of QCD in
presence of a finite density of baryonic matter.
In the present work we investigate the same issue for the case
of an external field, i.e. we will ascertain whether
deconfinement and chiral symmetry restoration do coincide
also in presence of a constant chromomagnetic field.

The paper is organized as follows.
In Section 2 we review our method to treat a background field on the lattice.
In Sections 3 and 4 we discuss numerical results and finally, in Section 5, we
present our conclusions.

\section{External fields on the lattice}
\label{extfields}

In this section we   review our method to study the dynamics
of lattice gauge theories in presence of background fields. In particular,
we focus on the case of  constant chromomagnetic fields.

\subsection{The method}
\label{themethod}

In Refs.~\cite{Cea:1997ff,Cea:1999gn} we introduced a lattice gauge invariant
effective action $\Gamma[\vec{A}^{\text{ext}}]$ for an external background
field $\vec{A}^{\text{ext}}$:
\be
\label{Gamma}
\Gamma[\vec{A}^{\text{ext}}] = -\frac{1}{L_t} \ln
\left\{
\frac{{\mathcal{Z}}[\vec{A}^{\text{ext}}]}{{\mathcal{Z}}[0]}
\right\}
\ee
where $L_t$ is the lattice size in time direction and
$\vec{A}^{\text{ext}}(\vec{x})$ is the continuum gauge potential of the
external static background field.  ${\mathcal{Z}}[\vec{A}^{\text{ext}}]$ is the
lattice functional integral
\be \label{Zetalatt}
{\mathcal{Z}}[\vec{A}^{\text{ext}}] =
\int_{U_k(\vec{x},x_t=0)=U_k^{\text{ext}}(\vec{x})} {\mathcal{D}}U \; e^{-S_W}
\,, \ee
with $S_W$ the standard pure gauge Wilson action.
The functional integration is performed over the lattice links, but constraining
the spatial links belonging to a given time slice (say $x_t=0$) to be
\be
\label{coldwall}
U_k(\vec{x},x_t=0) = U^{\text{ext}}_k(\vec{x})
\,,\,\,\,\,\, (k=1,2,3) \,\,,
\ee
 $U^{\text{ext}}_k(\vec{x})$ being the elementary parallel transports
corresponding to the external continuum
gauge potential $\vec{A}^{\text{ext}}(x)=\vec{A}^{\text{ext}}_a(x) \lambda_a/2$.
Note that the temporal links are not constrained.
${\mathcal{Z}}[0]$ is defined analogously, but adopting a zero
external field, i.e. with $U^{\text{ext}}_k(\vec{x})$ fixed to the
identity element of the gauge group.

In the case of a static background field which does not vanish at infinity we
must also impose that, for each time slice $x_t \ne 0$, spatial links exiting
from sites belonging to the spatial boundaries  are fixed according to
eq.~(\ref{coldwall}). In the continuum this last condition amounts to the
requirement that fluctuations over the background field vanish at infinity.

The partition function defined in eq.~(\ref{Zetalatt}) is also known
as lattice
Schr\"odinger functional~\cite{Luscher:1992an,Luscher:1995vs} and in the
continuum corresponds to the Feynman kernel~\cite{Rossi:1980jf}. Note that, at
variance with the usual formulation of the lattice Schr\"odinger
functional~\cite{Luscher:1992an,Luscher:1995vs}, where a cylindrical
geometry is adopted, our lattice has an hypertoroidal geometry, i.e.
the first and the last time slice are identified and periodic boundary
conditions are assumed in the time direction, so that the constraint
given in eq.~(\ref{Zetalatt}) should actually read
$U_k(\vec{x},L_t)=U_k(\vec{x},0)=U^{\text{ext}}_k(\vec{x})$.
With this prescription, $S_W$
in eq.~(\ref{Zetalatt}) is allowed to be the standard Wilson action.

The lattice effective action $\Gamma[\vec{A}^{\text{ext}}]$ defined
by eq.~(\ref{Gamma}) is given in terms of the
lattice Schr\"odinger functional, which is invariant for time-independent gauge
transformation of the background
field~\cite{Luscher:1992an,Luscher:1995vs}, therefore it is gauge
invariant too. It corresponds to the vacuum
energy, $E_0[\vec{A}^{\text{ext}}]$,
in presence of the background field, measured with respect to
the vacuum energy,
$E_0[0]$,
with  $\vec{A}^{\text{ext}}=0$
\be
\label{vacuumenergy}
\Gamma[\vec{A}^{\text{ext}}] \quad \longrightarrow \quad E_0[\vec{A}^{\text{ext}}]-E_0[0] \,.
\ee
The relation above is true by letting the temporal lattice size $L_t \to \infty$;
on finite lattices this amounts to have $L_t$ sufficiently large
to single out the ground state contribution to the energy.

For finite values of $L_t$, however, having adopted the prescription
of periodic boundary conditions in time direction, the functional
integral in eq.~(\ref{Zetalatt}) can be naturally interpreted as the
thermal partition function
${\mathcal{Z_T}}[\vec{A}^{\text{ext}}]$~\cite{Gross:1981br}
in presence of the background field $\vec{A}^{\text{ext}}$, with
the temperature given by $T=1/(a L_t)$.

In this case the  gauge invariant
effective action in eq.~(\ref{Gamma}) is replaced by the
the free energy functional defined as
\be
\label{freeenergy}
{\mathcal{F}}[\vec{A}^{\text{ext}}] = -\frac{1}{L_t} \ln
\left\{
\frac{{\mathcal{Z_T}}[\vec{A}^{\text{ext}}]}{{\mathcal{Z_T}}[0]}
\right\} \; .
\ee
When the physical temperature is sent to zero
the free energy  functional reduces to the vacuum energy functional,
eq.~(\ref{Gamma}).

Let us now consider the extension of the above formalism to full QCD,
i.e. including dynamical fermions, which is relevant for the present
work. In presence of dynamical fermions
the thermal partition functional
becomes~\cite{Cea:2004ux}
\begin{eqnarray}
\label{ZetaT}
\mathcal{Z}_T \left[ \vec{A}^{\text{ext}} \right]  &=
&\int_{U_k(L_t,\vec{x})=U_k(0,\vec{x})=U^{\text{ext}}_k(\vec{x})}
\mathcal{D}U \,  {\mathcal{D}} \psi  \, {\mathcal{D}} \bar{\psi} e^{-(S_W+S_F)}
\nonumber \\&=&  \int_{U_k(L_t,\vec{x})=U_k(0,\vec{x})=U^{\text{ext}}_k(\vec{x})}
\mathcal{D}U e^{-S_W} \, \det M \,,
\end{eqnarray}
where $S_F$ is the fermion action and $M$ is the fermionic matrix.
The spatial links  are still constrained to values corresponding
to the external background field, whereas
the fermionic fields are not constrained.
The relevant quantity is still the free energy functional defined as
in eq.~(\ref{freeenergy}).

Actually, a direct numerical evaluation of the ratio of partition
functions appearing in eq.~(\ref{freeenergy}) turns out to be quite
difficult. Even if techniques have been developed recently to deal
with similar problems~\cite{deForcrand:2000fi,D'Elia:2006vg}, we adopt the more conventional
strategy~\cite{DelDebbio:1994sx,Cea:1997ff}
of computing instead
a susceptibility of the free energy functional, in particular its derivative
$F^\prime$
with respect to the inverse gauge coupling $\beta$, which can be
easily evaluated and is also more appropriate for the aim of the
present study. $F^\prime$ is defined as
\begin{equation}
\label{deriv}
F^\prime(\beta) =
\frac{\partial {\mathcal{F}}(\beta)}{\partial \beta} =
\left \langle
 \sum_{x,\mu < \nu}
\frac{1}{3} \,  \text{Re}\, {\text{Tr}}\, U_{\mu\nu}(x) \right\rangle_0  \\
  - \left\langle  \sum_{x,\mu< \nu} \frac{1}{3} \,  \text{Re} \, {\text{Tr}} \, U_{\mu\nu}(x)
\right\rangle_{\vec{A}^{\text{ext}}} \,,
\end{equation}
where the subscripts on the averages indicate the value of the
external field.
Only unconstrained plaquette are taken into account in the sum
in eq.~(\ref{deriv}).
Observing that $F[\vec{A}^{\text{ext}}] = 0$ at $ \beta = 0$, we
may eventually obtain  $F[\vec{A}^{\text{ext}}]$ from
$F^{\prime}[\vec{A}^{\text{ext}}]$ by numerical integration:
\be
\label{trapezu1}
F[\vec{A}^{\text{ext}}]  =  \int_0^\beta
F^{\prime}[\vec{A}^{\text{ext}}] \,d\beta^{\prime} \; .
\ee

\subsection{A constant chromomagnetic field on the lattice}
\label{constantfield}

Let us now define a static constant abelian chromomagnetic field on the lattice.
In the continuum the gauge potential giving rise to a static constant abelian chromomagnetic field
directed along spatial direction $\hat{3}$ and direction $\tilde{a}$
in the color space can be written in the following form:
\be
\label{su3pot}
\vec{A}^{\text{ext}}_a(\vec{x}) =
\vec{A}^{\text{ext}}(\vec{x}) \delta_{a,\tilde{a}} \,, \quad
A^{\text{ext}}_k(\vec{x}) =  \delta_{k,2} x_1 H \,.
\ee
In SU(3) lattice gauge theory
the constrained lattice links (see eq.~(\ref{coldwall})) corresponding to
the continuum gauge potential eq.~(\ref{su3pot}) are (choosing $\tilde{a}=3$, i.e. abelian
chromomagnetic field along direction $\hat{3}$ in color space)
\be
\label{t3links}
\begin{split}
& U^{\text{ext}}_1(\vec{x}) =
U^{\text{ext}}_3(\vec{x}) = {\mathbf{1}} \,,
\\
& U^{\text{ext}}_2(\vec{x}) =
\begin{bmatrix}
\exp(i \frac {a g H x_1} {2})  & 0 & 0 \\ 0 &  \exp(- i \frac {a g H
x_1} {2}) & 0
\\ 0 & 0 & 1
\end{bmatrix}
\,.
\end{split}
\ee
We will refer to this case as $T_3$ abelian chromomagnetic field,
which will be our choice in the present work. Of course it is possible
to choose various alternatives, like an abelian field along
the direction of the $T_8$ generator, or along different combinations
of $T_3$ and $T_8$.

Since our lattice has the topology of a torus,
the magnetic field turns out to be quantized
\be
\label{quant} a^2 \frac{g H}{2} = \frac{2 \pi}{L_1}
n_{\text{ext}} \,, \qquad  n_{\text{ext}}\,\,\,{\text{integer}}\,.
\ee
In the following $n_{\text{ext}}$ will be used to parameterize the
external field strength.

\section{Numerical simulations and results}
\label{numericalresults}

We have studied full QCD dynamics with two flavors of staggered
fermions in presence of a constant
chromomagnetic field. The  simulations have been performed
on lattices $32^3\times8$ and $64\times32^2\times8$.
We used a slight modification of the standard HMC R-algorithm~\cite{Gottlieb:1987mq}
for two degenerate flavors of
staggered fermions with quark mass
$a m_q = 0.075$. According to our previous discussion,  the links which are frozen
are not evolved during the molecular dynamics trajectory and the corresponding conjugate
momenta are set to zero.
We have collected about 2000
thermalized trajectories for each value of $\beta$.
Each trajectory consists of $125$  molecular dynamics
steps and has total length  $1$. The computer simulations have
been performed using computer facilities at the
INFN apeNEXT computing center in Rome.

\subsection{The critical coupling}

As is well known,
the pure SU(3) gauge system undergoes a deconfinement
phase transition at a given critical temperature and this happens even in the unquenched
case (see Ref.~\cite{Laermann:2003cv} for an up-to-date review).
In our earlier studies~\cite{Cea:2002wx,Cea:2005td} we found that the critical coupling in pure
non-abelian gauge theories is shifted towards lower values by immersing the system in a constant chromomagnetic
background field: that means lower temperatures on lattices where the
temporal extent in lattice units is kept constant ($T = 1/(L_t a)$).
The main purpose of the present study is to verify if this effect survives
in presence of dynamical fermions. We refer in the
following to the constant abelian background field
defined in Eqs.~(\ref{su3pot}) and (\ref{t3links}).

In order to evaluate the critical gauge coupling we measure
$F^\prime[\vec{A}^{\text{ext}}]$ (eq.~(\ref{deriv})),
the derivative of the free energy with respect to the gauge coupling $\beta$, as a function of $\beta$.
We found that $F^\prime[\vec{A}^{\text{ext}}]$ displays
a peak in the critical region
where it can be parameterized as
\be
\label{peak-form}
F^{\prime}(\beta,L_t)
= \frac{a_1(L_t)}{a_2(L_t) [\beta - \beta^*(L_t)]^2 +1} \,.
\ee
In figure 1
%
%
\FIGURE[ht]{\label{Fig1}
\includegraphics[width=0.85\textwidth,clip]{figure_1.eps}
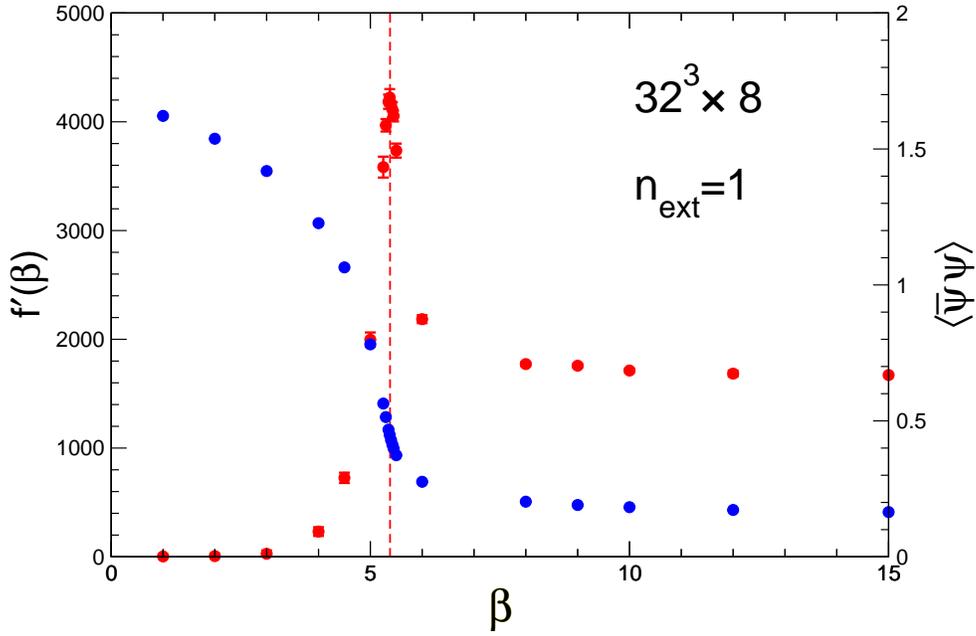
\caption{The derivative of the free energy eq.~(\ref{deriv}) with
respect to the gauge coupling (left axis, blue circles), and the
chiral condensate eq.~(\ref{chiralcond}) (right axis, red squares)
versus $\beta$. The vertical line represents the position of the
peak in the derivative of the free energy.}
}
we show an example of $F^\prime$ measured for
$n_{\text{ext}} = 1$ on a $32^3 \times 8$ lattice. In the same figure
we display also the chiral condensate
\be
\label{chiralcond}
\langle \bar{\psi} \psi \rangle = \langle \frac{1}{V} \, \frac{N_f}{4} \,\, \text{Tr}\, M^{-1} \rangle
\ee
and the numerical data point out that the peak in the derivative of
the free energy corresponds to the drop in the chiral condensate,
the latter is a signal of the transition leading to chiral symmetry
restoration.

In figure~\ref{Fig2}
%
%
\FIGURE[ht]{\label{Fig2}
\includegraphics[width=0.85\textwidth,clip]{figure_2.eps}
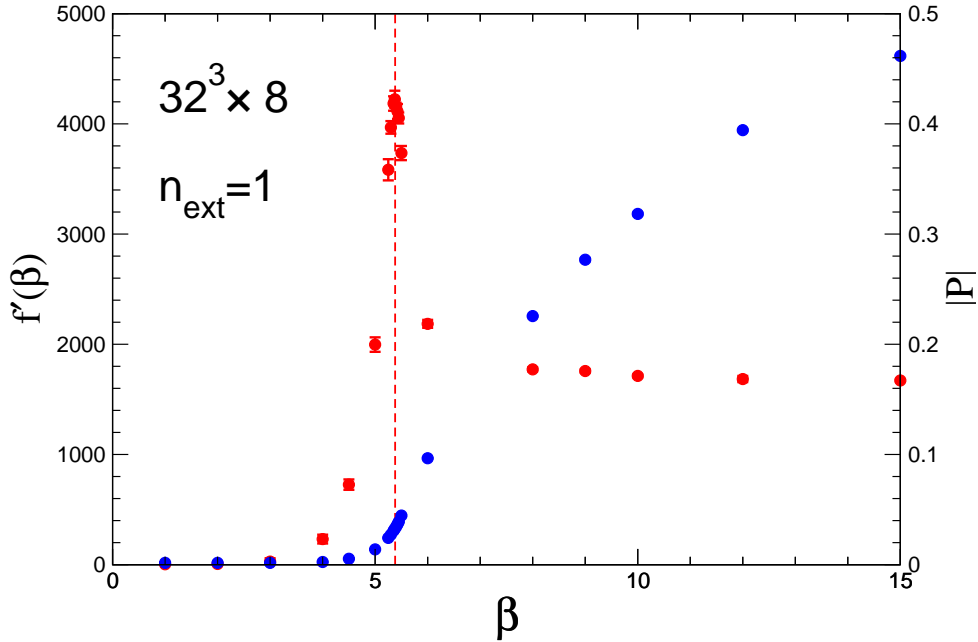
\caption{The derivative of the free energy as in figure~\ref{Fig1}
together with the absolute value of the Polyakov loop eq.~(\ref{Polyakov}).
Vertical dotted line as in figure~\ref{Fig1}.}
}
we compare the derivative of the free energy  with
the absolute value of the Polyalov loop
\be
\label{Polyakov}
P = \frac{1}{V_s} \sum_{\vec{x}} \frac{1}{3} \, {\text{Tr}}  \prod_{x_4=1}^{L_t} U_4(x_4,\vec{x}) \; .
\ee
We can see that also in this case the rise of the Polyakov loop
(expected at the
deconfining phase transition) corresponds
to the peak of the derivative of the free energy.
Moreover, in figure~\ref{Fig3}
%
%
\FIGURE[ht]{\label{Fig3}
\includegraphics[width=0.85\textwidth,clip]{figure_3.eps}
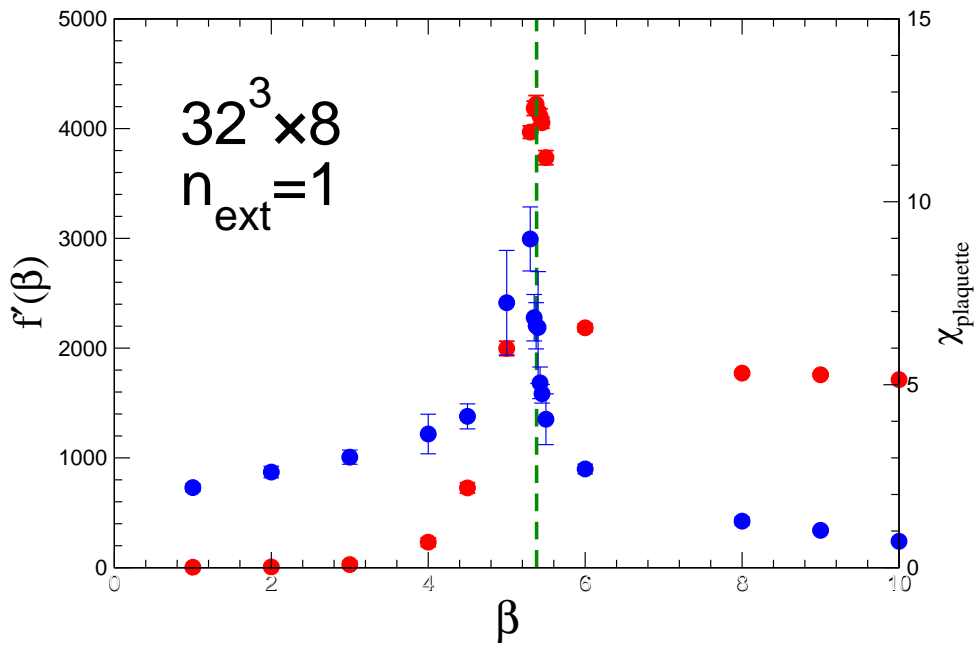
\caption{The derivative of the free energy as in figure~\ref{Fig1}
together with the susceptibility of the gauge action.}
}
the derivative of the free energy is
displayed together with the plaquette susceptibility
(susceptibility of the gauge action).
It is evident from this figure that, within our statistical
uncertainties, the peaks of the two quantities coincide.
Similar results are obtained by looking at the susceptibilities
of the Polyakov loop and of the chiral condensate.

From the above arguments we may draw some partial conclusions:
the critical
coupling of the phase transition can be located by looking at the peak
of the derivative of the free energy; moreover, as in the case
of zero external field and within statistical uncertainties,
 a single transition seems to be present
where both deconfinement and chiral symmetry restoration take place.

It is worth to note that since the measurement of the derivative of
the free energy is simply related to the measurement of the gauge
plaquette we may have a good evaluation of the critical coupling
with a relatively small sample of measurements.

As told before our aim is to find if the critical coupling depends on the strength of the applied constant
chromomagnetic field. To this purpose we have varied the strength of
the external field by tuning up the parameter $n_{\text{ext}}$
and we have searched for the phase transition signalled by the
peak of the derivative of the free energy.
We have found that indeed the critical coupling shifts towards lower
values  by increasing the external field strength.
In figure~\ref{Fig4}
%
%
\FIGURE[ht]{\label{Fig4}
\includegraphics[width=0.85\textwidth,clip]{figure_4.eps}
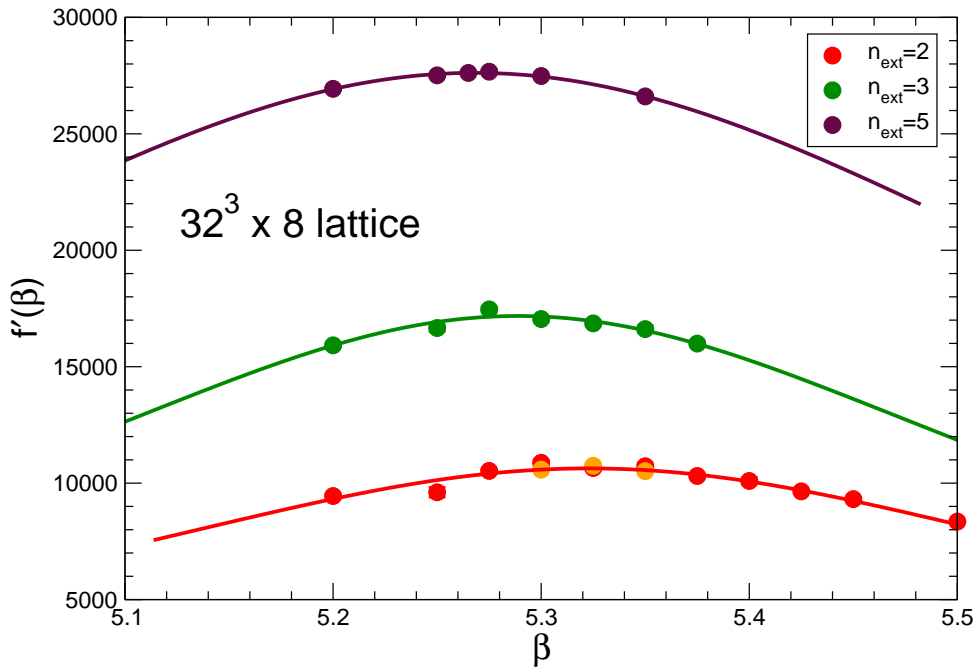
\caption{The derivative of the free energy eq.~(\ref{deriv}) versus $\beta$
for some values of the strength of the constant chromomagnetic field
parameterized (see eq.~(\ref{quant})) by the integer $n_{\text{ext}}$.
The yellow full circles corresponds to runs performed with
different machines (APEmille) and algorithms as a check.}
}
we display the derivative of the free energy in
correspondence of three values of $n_{\text{ext}}$
obtained on a $32^3\times8$ lattice,
together with the fit curves given by eq.~(\ref{peak-form}).
As one can see the position of the peaks decreases by increasing the external field strength.
In Table~1
\TABLE[t]{
\setlength{\tabcolsep}{0.9pc}
\centering
\caption[]{The values of the critical coupling versus the external field strengths.}
\begin{tabular}{ccc}
\hline
\hline
\multicolumn{1}{c}{lattice size}
& \multicolumn{1}{c}{$n_{\text{ext}}$}
& \multicolumn{1}{c}{$\beta_c$} \\
\hline
$\qquad 32 \times 32 \times 32 \times 8 \qquad $    & $\qquad 0 \qquad$    & 5.4851 (202) \\
$\qquad 64 \times 32 \times 32 \times 8 \qquad $    & $\qquad 1 \qquad$    & 5.4288 (128)\\
$\qquad 32 \times 32 \times 32 \times 8 \qquad $    & $\qquad 1 \qquad$    & 5.3808 (128)\\
$\qquad 32 \times 32 \times 32 \times 8 \qquad $    & $\qquad 2 \qquad$    & 5.3228 (90)\\
$\qquad 32 \times 32 \times 32 \times 8 \qquad $    & $\qquad 3 \qquad$    & 5.2888 (44)\\
$\qquad 32 \times 32 \times 32 \times 8 \qquad $    & $\qquad 4 \qquad$    & 5.2659 (48)\\
$\qquad 32 \times 32 \times 32 \times 8 \qquad $    & $\qquad 5 \qquad$    & 5.2680 (38)\\
\hline
\hline
\end{tabular}
\label{table1}
}
we report the values of the critical couplings versus
$n_{\text{ext}}$.

Figure~\ref{Fig5}
%
%
\FIGURE[ht]{\label{Fig5}
\includegraphics[width=0.85\textwidth,clip]{figure_5.eps}
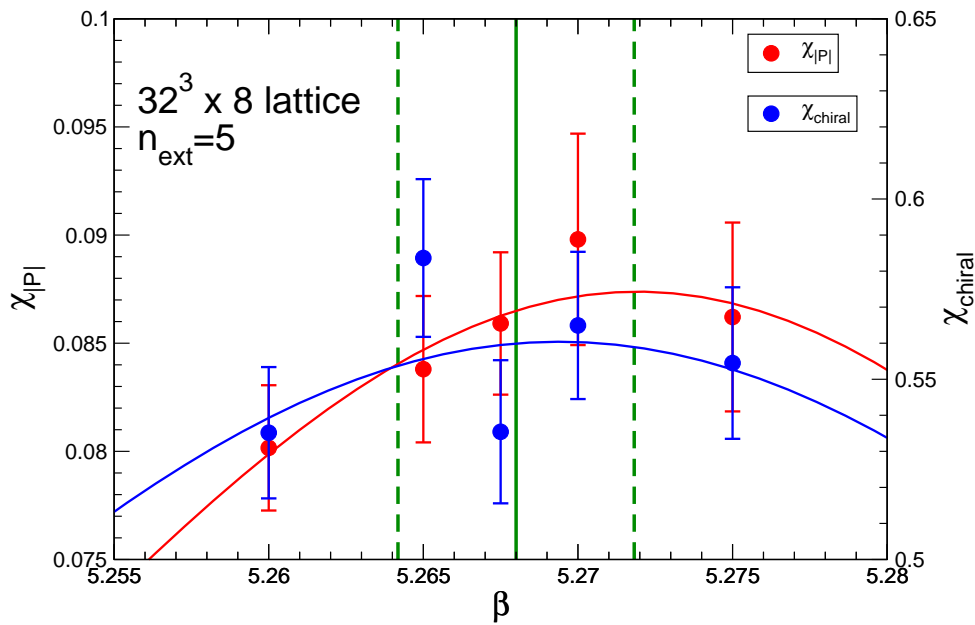
\caption{The susceptibility of the absolute value of the Polyakov loop together with
the susceptibility of the chiral condensate. The vertical full line represents
the position of the peak in the derivative of the free energy for chromomagnetic
field strength $n_{\text{ext}}=5$. The vertical dotted lines give the error region.
Red and blue full lines are the best fits with the same parameterization as in eq.~(\ref{peak-form}) respectively to
the susceptibility of the (absolute value of the) Polyakov loop and
to that of the chiral condensate.}
}
displays instead the susceptibility of the
absolute value of the Polyakov loop together with the
susceptibility of the chiral condensate in the peak region for
the largest explored value of the external magnetic field
($n_{\text{ext}} = 5$).
It is worthwhile to note that, as mentioned earlier,
the position of the peak of Polyakov loop  susceptibility ($\beta=5.2719(164)$)
and the position of the peak of the chiral condensate susceptibility
($\beta=5.2694(84)$) are consistent with each other
and with the position of the peak obtained from the derivative of the
free energy ($\beta=5.2680(38)$), thus confirming the conclusion
stated above, i.e. that the chiral and the deconfinement
transition are shifted towards lower temperatures by the presence
of the external field in an equal way, i.e. they continue
to be coincident within statistical errors even for $n_{\text{ext}} \neq 0$.

The numerical results obtained so far let us conclude that the critical coupling is dependent
on the strength of the background constant chromomagnetic field. On the other hand for pure SU(3) gauge theory we
obtained~\cite{Cea:2003un} that the value of the critical coupling is not changed by a monopole
background field.
Similar results are expected in presence of
dynamical fermions~\cite{Carmona:2002ty,Cea:2004ux,D'Elia:2005ta},
however we will verify this fact explicitly for the present case.

We recall the definition
of an abelian monopole background field on the lattice, for more
details and physical results see ref.~\cite{Cea:2004ux}.
It is well known that for SU(3) gauge theory the maximal abelian group is
U(1)$\times$U(1), therefore we may introduce two independent types
of abelian monopoles using respectively the Gell-Mann matrices
$\lambda_3$ and $\lambda_8$ or their linear combinations.
In the following we shall consider the abelian monopole field related to
the $\lambda_3$ diagonal generator.
In the continuum the abelian monopole field is given by
\begin{equation}
\label{monop3su3}
g \vec{b}^a({\vec{x}}) = \delta^{a,3} \frac{n_{\mathrm{mon}}}{2}
\frac{ \vec{x} \times \vec{n}}{|\vec{x}|(|\vec{x}| -
\vec{x}\cdot\vec{n})} \,,
\end{equation}
where $\vec{n}$ is the direction of the Dirac string and,
according to the Dirac quantization condition, $n_{\text{mon}}$ is
an integer. The lattice links corresponding to the abelian
monopole field eq.~(\ref{monop3su3}) are (we choose $\vec{n}=\hat{x}_3$)
\begin{equation}
\label{t3linkssu3}
\begin{split}
U_{1,2}^{\text{ext}}(\vec{x}) & =
\begin{bmatrix}
e^{i \theta^{\text{mon}}_{1,2}(\vec{x})} & 0 & 0 \\ 0 &  e^{- i
\theta^{\text{mon}}_{1,2}(\vec{x})} & 0 \\ 0 & 0 & 1
\end{bmatrix}
\,  \\ U^{\text{ext}}_{3}(\vec{x}) & = {\mathbf 1} \,,
\end{split}
\end{equation}
with $\theta^{\text{mon}}_{1,2}(\vec{x})$ defined as
\begin{equation}
\label{thetat3su2}
\begin{split}
\theta^{\text{mon}}_1(\vec{x}) & = -\frac{a n_{\text{mon}}}{4}
\frac{(x_2-X_2)}{|\vec{x}_{\text{mon}}|}
\frac{1}{|\vec{x}_{\text{mon}}| - (x_3-X_3)} \,, \\
\theta^{\text{mon}}_2(\vec{x}) & = +\frac{a n_{\text{mon}}}{4}
\frac{(x_1-X_1)}{|\vec{x}_{\text{mon}}|}
\frac{1}{|\vec{x}_{\text{mon}}| - (x_3-X_3)} \,,
\end{split}
\end{equation}
where  $(X_1,X_2,X_3)$ are the monopole coordinates,
$\vec{x}_{\text{mon}} = (\vec{x} - \vec{X})$.
The monopole background field is introduced
by constraining (see eq.~(\ref{t3linkssu3})) the spatial links exiting from the sites
at the boundary of the time slice $x_t=0$. For what concern spatial links exiting from sites
at the boundary of other time slices ($x_t \ne 0$) we  constrain these links according to eq.~(\ref{t3linkssu3}).

We have performed numerical simulations
in presence of an abelian monopole background field
with monopole charge $n_{\text{mon}}=10$
(again for 2 staggered flavors QCD of mass $a m_q=0.075$).
The critical coupling has been located by looking at the
%
%
\FIGURE[ht]{\label{Fig6}
\includegraphics[width=0.85\textwidth,clip]{figure_6.eps}
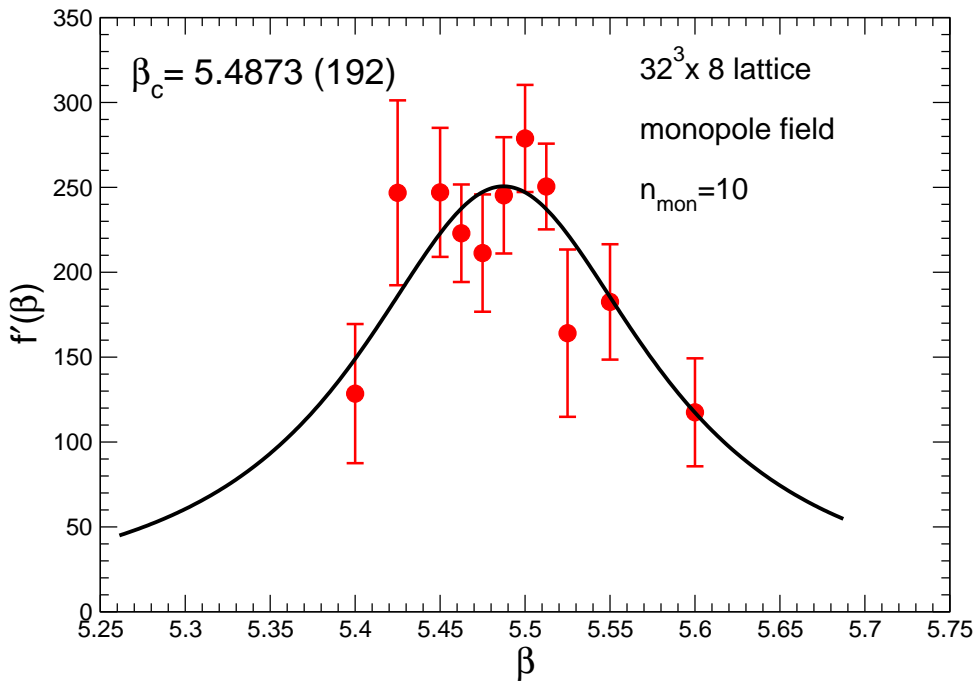
\caption{The derivative of the free energy for the abelian monopole background field
eq.~(\ref{monop3su3} in the peak region, together with (full line) the best fit
eq.~(\ref{peak-form}).}
}
peak of the derivative of the free energy (see figure~(\ref{Fig6})). We find
\be
\label{monopole_peak}
\beta_c = 5.4873(192) \,.
\ee
We have also done simulations
in absence of any external chromomagnetic field,
finding that the susceptibilities of the Polyakov loop,
of the chiral condensate and of the plaquette display
a peak at
\be
\label{beta_c_wth_f}
\beta_c = 5.495 (25) \;.
\ee
Noticeably, this value of the critical coupling without external field is consistent, within our statistical uncertainty, with the value we get
when we consider an abelian monopole field as background field.
Therefore we can conclude that, as we found~\cite{Cea:2004ux} in the case of pure gauge SU(3) gauge theory,
the abelian monopole field has no effect on the position of the critical coupling.

\section{Deconfinement temperature and critical field strength}

In previous studies~\cite{Cea:2005td} in  pure lattice gauge
theories we looked for  the possible dependence of the deconfinement
temperature on the strength of an external (chromo)magnetic field.
In particular we studied  SU(2), and SU(3) l.g.t.'s both in (2+1)
and (3+1) dimensions and U(1) l.g.t. both in 4 dimensions and in
(2+1) dimensions. In fact, in the case of non-abelian gauge
theories, irrespective of the number of dimensions, we found that
the deconfinement temperature depends on the strength of the
constant chromomagnetic background field
(similar studies have been performed within a different framework
in refs.~\cite{Skalozub:1999bf,Demchik:2006qj}).
On the other hand, for  U(1) gauge theory
we found no evidence for a dependence of
the critical coupling on the strength of an external magnetic
field. In particular, as is well known,   4-dimensional U(1) l.g.t.
undergoes a transition from a confined phase to a Coulomb phase: our
analysis showed~\cite{Cea:2005td} that the location of the
confinement-Coulomb phase transition is not changed by varying the
strength of an applied constant magnetic field. The same analysis has been performed for
compact quantum electrodynamics in (2+1) dimensions where it is
known~\cite{Polyakov:1976fu} that at zero temperature external
charges are confined for all values of the coupling and it is well
ascertained that the confining mechanism is the condensation of
magnetic monopoles which gives rise to a linear confining potential
and a non-zero string tension. Even in this case we verified that
the critical temperature for deconfinement does not depend on the
strength of an external magnetic field. As a consequence, we concluded that
the dependence of the critical coupling on the strength of the
external chromomagnetic field is a peculiar feature of non-abelian
theories.

The main aim of the present investigation is to
extend our study to non-abelian gauge theories in presence
of dynamical fermions.
To this end, having determined in the previous section the critical couplings corresponding to different external field strengths, we now try to estimate the critical temperature:
\be
\label{criticaltemp}
T_c = \frac{1}{a(\beta_c, m_q) L_t}  \,,
\ee
where $L_t$ is the lattice temporal size and $a(\beta_c, m_q)$
is the lattice spacing at the given critical coupling $\beta_c$.
In the case of SU(3) pure gauge theory in order to evaluate $a(\beta_c)$ we used~\cite{Cea:2005td}  the string tension.
We obtained that the values of the critical temperature versus the
square root of the external field strength can be fitted
by a linear parameterization.  By extrapolating to zero external field strength we obtained:
\be
\label{Tczerofield}
\frac{T_c(0)}{\sqrt{\sigma}} = 0.643(15)
\ee
in very good agreement with the estimate $T_c/\sqrt{\sigma}=0.640(15)$ in the literature~\cite{Teper:1998kw}.
At the same time the intercept of  line with the zero temperature axis furnished an
estimate of the critical field strength (i.e. the limit value above which the
gauge system is in the deconfined phase even at very low temperatures)
\be
\label{criticalfield}
\sqrt{gH_c} = (2.63  \pm 0.15) \sqrt{\sigma} = (1.104 \pm 0.063) {\text{ GeV}} = 6.26(2) \times 10^{19} \text{ Gauss}
\ee
using for the physical value of the string tension $\sqrt{\sigma}=420 \text{ MeV}$.
The same analysis can be performed by means of the improved lattice scale introduced
in Refs.~\cite{Allton:1996kr,Edwards:1998xf}
\be
\label{lambdaimproved}
\widetilde{\Lambda}=\frac{1}{a} f(g^2) (1+c_2 \hat{a}(g)^2 + c_4 \hat{a}(g)^4) \,, \;\; \hat{a}(g)\equiv \frac{f(g^2)}{f(g^2=1)}
\ee
where  $g$ is the gauge coupling, $c_2=0.195(16)$, $c_4=0.0562(45)$, $\widetilde{\Lambda}/\sqrt{\sigma}=0.0138(12)$
and $f(g^2)$ is the 2-loop scaling function
\be
\label{asympscaling}
f(g^2) = (b_0 g^2)^{-b_1/2b_0^2}  \,\, \exp\left(- \frac{1}{2 b_0 g^2} \right)
\ee
with
\be
\label{coeffs}
\begin{split}
&b_0 = \frac{1}{16 \pi^2} \left[ 11 \frac{N_c}{3} - \frac{2}{3} N_f \right] \\
&b_1 = \left(\frac{1}{16 \pi^2} \right)^2 \left[ \frac{34}{3} N_c^2 -
    \left( \frac{10}{3} N_c + \frac{N_c^2 -1}{N_c} \right) N_f \right]
  \, ;
\end{split}
\ee
$N_c$ is the number of colors and $N_f$ is the number of flavors.

Obviously the analysis of pure gauge data in units of $\widetilde{\Lambda}$ gives results consistent with
the same analysis done using the scale of the string tension (see Ref.~\cite{Cea:2005td}). In particular
a linear extrapolation towards zero external field gives:
\be
\label{estrapsu3}
\frac{T_c}{\widetilde{\Lambda}} = 45.05 (1.02)
\ee
in agreement with $T_c/\widetilde{\Lambda}=46.38(1.09)$ obtained from
$T_c/\sqrt{\sigma}=0.640(15)$.
Moreover the critical field strength turns out to be
\be
\label{criticalfield2}
\frac{\sqrt{gH_c}}{\widetilde{\Lambda}} = 209.6 \pm 3.07
\ee
which corresponds to $\sqrt{gH_c}=1.21(11) GeV$ in agreement with the estimate $\sqrt{gH_c}=1.10(6) GeV$
given in eq.~(\ref{criticalfield}).

Let us turn now to the $N_f=2$ case. Also in this case we have to face the
problem of fixing the physical scale.
In order to reduce the systematic effects involved in this procedure
we will consider the ratios
\be
\label{ratios}
\frac{T_c(gH)}{T_c}  \qquad \text{vs.} \qquad \frac{\sqrt{gH}}{T_c}
\ee
where $T_c$ is the critical temperature without external field.
The above quantities can be obtained once
the ratio of the lattice spacings at the respective couplings is
known. A rough estimate of this ratio can be inferred by using the
2-loop scaling function $f(g^2)$ given in eq.~(\ref{asympscaling})
for $N_f = 2$. A better estimate could be obtained, as in the quenched
case, by adopting an improved scaling function
$f(g^2) (1+c_2 \hat{a}(g)^2 + c_4 \hat{a}(g)^4)$. We do not know, however,
the values of $c_2$ and $c_4$ for $N_f = 2$. In a first
approximation we will fix $c_2$ and $c_4$ to their quenched
values given above. In figure~\ref{Fig7}
%
%
%
\FIGURE[ht]{\label{Fig7}
\includegraphics[width=0.85\textwidth,clip]{figure_7.eps}
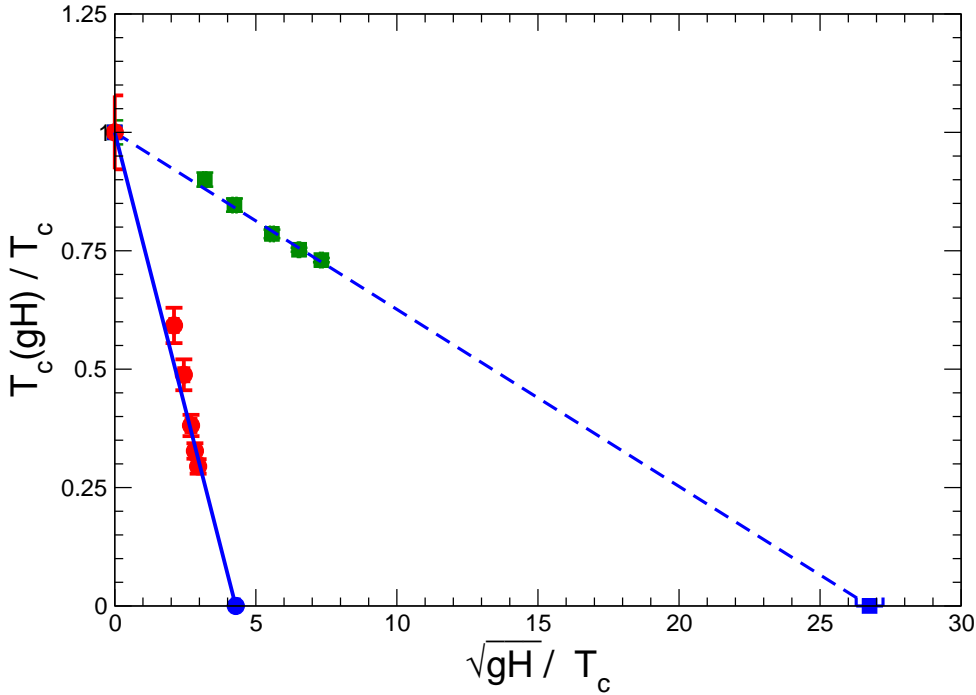
\caption{The critical temperature $T_c(gH)$ at a given strength of the chromomagnetic
background field in units of the critical temperature $T_c$ without external field versus
the square root of the strength of the background field in the same units.
Red circles are obtained by adopting the improved scaling function.
The blue line is the linear best fit. The blue circle  on the horizontal axis  is
the linear extrapolated values for the critical background field.
Green squares are obtained by adopting the 2-loop scaling function.
The blue dashed line is the linear best fit. The blue square  on the horizontal axis  is
the linear extrapolated values for the critical field.}
}
the quantities
reported in eq.~(\ref{ratios}) are displayed for both choices
described above, i.e. 2-loop asymptotic scaling and improved scaling.

The main result of our investigation is clear:  even in presence of
dynamical quarks the critical temperature decreases  with the strength
of the chromomagnetic field; moreover a linear fit to our data can be
extrapolated to very low temperatures, leading to the prediction of
a critical field strength above which strongly interacting matter
should be deconfined at all temperatures.

However, as can be appreciated from figure~\ref{Fig7}, the exact value
of the critical field strength is largely dependent on the choice
of the physical scale. In particular, assuming the 2-loop
scaling function we obtain
\be
\label{gHTcaf}
\frac{\sqrt{gH_c}}{T_c} = 26.8 (5) \,,
\ee
while assuming the improved scaling function we obtain
\be
\label{gHTc}
\frac{\sqrt{gH_c}}{T_c} = 4.29 (10) \,.
\ee
If the deconfinement temperature at zero field strength is taken
to be of the order of $170$ MeV, that means
$\sqrt{gH_c}$ in the range $0.7 - 4.5$ GeV.

It is clear that in order to get a reliable estimate of the critical
field strength, suitable for phenomenological purposes,
one should get a more reliable estimate of the physical
scale of our lattices. However one should work at a fixed value
of the pion mass (as close as possible to its physical value)
as well: that is beyond the purpose of the present investigation
and will be the subject of further studies.

To conclude this section we consider our measurements of the chiral condensate. In figure~\ref{Fig8}
%
%
\FIGURE[ht]{\label{Fig8}
\includegraphics[width=0.85\textwidth,clip]{figure_8.eps}
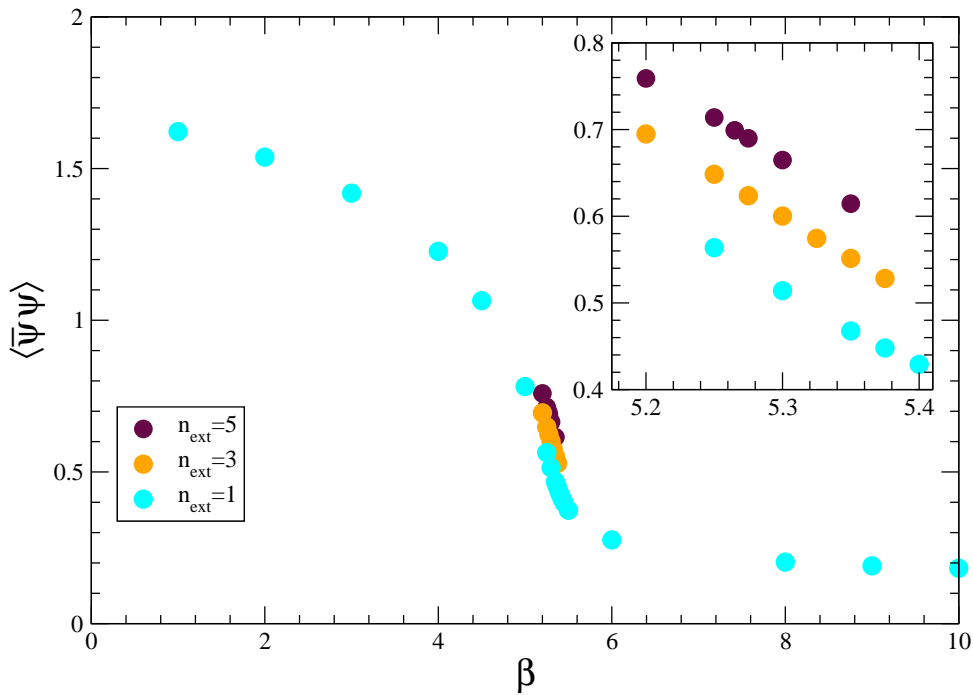
\caption{The chiral condensate eq.~(\ref{chiralcond}) versus $\beta$ in correspondence
of some values of the constant chromomagnetic background field.
In the inset the region corresponding to the phase transition has been magnified.}
}
we display the chiral condensate versus the gauge coupling in correspondence of some values of
the external field strength. Our numerical data show that, at least in the critical region, the value of the chiral condensate depends
on the strength of the applied field. Interestingly enough,
similar results for the chiral condensate have been found in ref.~\cite{Alexandre:2001pa}.

\section{Summary and Conclusions}
In this paper we have studied how a constant chromomagnetic field
perturbs the QCD dynamics.
In particular we focused on the theory at finite temperature
and we have found that, analogously to what happens in the pure
gauge theory~\cite{Cea:2005td}, the critical temperature
depends on the strength of the constant chromomagnetic background
field: it decreases as the external field is increased
and we have inferred,
as an extrapolation of our results,
that eventually the system is always deconfined for strong enough field
strengths. We estimated this critical field strength to be of the order
of 1~GeV, which is a typical QCD scale~\cite{Kabat:2002er}.
Notice that our estimate is of course affected by several systematic
 uncertainties, like that regarding the estimate of the physical
scale: for that reason we consider it only as an order of magnitude
of the real expected critical strength. In fact a more reliable determination
usable for phenomenological purposes should be performed by working
with a fixed physical value of the pion
mass (and as close as possible to its physical value); that
is out of the purpose of the present work and will be the subject
of future investigations.

By comparing the critical couplings determined from the
derivative of the free energy functional with those determined
from the susceptibility of the chiral condensate and of the Polyakov
loop we have ascertained that, even in presence of an external
chromomagnetic background field and at least
up to the field strengths explored in the present work,
the critical temperatures
where deconfinement and chiral symmetry restoration take place
coincide within errors.

Another intriguing aspect we have found is the dependence of the
chiral condensate on the chromomagnetic field strength.
This last point deserves further studies.
In order to get a deeper
understanding of our results, we also plan to study the effect of the
background field on the equation of state of QCD.

\providecommand{\href}[2]{#2}\begingroup\raggedright\endgroup

\end{document}